\documentclass[epj]{svjour}
\usepackage{epsfig}
\usepackage{graphicx}
\usepackage{graphics}
\usepackage{subfigure}
\usepackage{multirow}
\usepackage{hyperref}
\ifx\pdfoutput\undefined
\else
\pdfpagewidth=\paperwidth
\pdfpageheight=\paperheight
\fi 
\hypersetup{colorlinks=false,breaklinks=true,plainpages=false}

\begin{document}
\title{Vector bosons in heavy-ion collisions at the LHC}
\author{Zaida Conesa del Valle\inst{}
}                     % Do not remove
\authorrunning{Z. Conesa del Valle}
%
%\offprints{}          % Insert a name or remove this line
%
\institute{
Laboratoire Leprince-Ringuet (\'Ecole Polytechnique, CNRS-IN2P3),
Palaiseau, France%, {\it zconesa@in2p3.fr}}
}
\date{Received: August 28, 2008}
% The correct dates will be entered by Springer
%

\abstract{
Vector bosons become accessible experimental probes in heavy-ion collisions at the LHC. 
The capabilities of the LHC experiments to perform their measurement are outlined. 
The focus is given to their utility to study the possible formation and properties of the Quark Gluon Plasma (QGP) in the most central heavy-ion collisions. 
Their own sensitivity (if any) to the QGP is discussed. 
Their interest as references to observe multiple QGP sensitive probes is justified. 
 % end of PACS codes
} %end of abstract

\maketitle

\section{Introduction}
\label{intro}

W$^{\pm}$ and Z$^0$ bosons are massive weakly interacting probes. They have been extensively studied at CERN, SLAC and FNAL in $p \bar{p}$ and $e^+ e^-$ collisions~\cite{pdg} and are considered as standard model benchmarks. 
At the Large Hadron Collider (LHC), about 3 W (Z) are expected to decay in leptonic channels per each 10 (100) million pp collisions\footnote{
Calculations are done with the values of Tab.~\ref{tab:cross_section} and considering $\sigma^{inel}=70$~mb ($8$~b) in pp (Pb-Pb) collisions.
} at $\sqrt{s_{NN} }=14$~TeV, and up to 30 W (Z) per 1 (10) million Pb-Pb collisions at $5.5$~TeV.
A large enough amount that will provide the opportunity to observe them for the first time in heavy-ion collisions (HIC) at the LHC. 
\newline
\begin{table}
\caption{W and Z production cross-sections per nucleon-nucleon collision from NLO calculations. Shadowing is included in Pb-Pb calculations.}
\label{tab:cross_section}   
\begin{tabular}{cccc}
\hline\noalign{\smallskip}
 collision ($\sqrt{s_{NN}} \, [TeV]$)& pp (14) & Pb-Pb (5.5) 
 \\ \noalign{\smallskip}\hline\noalign{\smallskip}
 $\sigma_{NN}^{W} \, \times BR_{\mu \nu}$ [nb]    & 20.9~\cite{frixione_mangano} & 6.56~\cite{vogt,vogt_talk} 
 \\ \noalign{\smallskip}
 $\sigma_{NN}^{Z} \, \times BR_{\mu^+ \mu^-}$ [nb] & 1.9~\cite{tricoli}  &  0.63~\cite{vogt,vogt_talk} 
 \\ \noalign{\smallskip}\hline
\end{tabular}
\end{table}
Weak bosons at the LHC will be multi-purpose observables. 
More than $80\%$ of their total production cross-section is accounted for by quark anti-quark scattering processes, whereas NLO processes amount to about $13\%$ and NNLO corrections are just of $1$-$2\%$~\cite{anastasiou,frixione_mangano}. 
Consequently, they are considered as benchmarks, they have been suggested as 'standard-candles' for luminosity measurements and as tools to ameliorate the detectors knowledge~\cite{martin}. 
In addition, they will be sensitive to the quarks PDFs and to their nuclear modifications at $Q^2  \approx M^{2}_{W(Z)}$. 
In this paper we focus on their production in heavy-ion collisions and, in particular, on their interest to study the formation and properties of the Quark Gluon Plasma (QGP). 
Vector bosons do not interact strongly, but is it a sufficient condition to consider them as medium-blind references? To understand their role in HIC it becomes imperative to discuss the influence, if any, of this hot and dense medium on them and/or on their decay products. 
\newline
A qualitative estimate of the formation and decay time of vector bosons with respect to the QGP formation and evolution will facilitate the comprehension of the physics processes involved. 
Weak bosons are formed early due to their large mass $\sim 1/M \sim 10^{-3}$~fm/c. Their decay time is by definition inversely proportional to their widths, that is $0.08$~fm/c for the Z and $0.09$~fm/c for the W~\cite{pdg}. 
In the most extreme approach, the QGP would thermalise immediately and the weak bosons would decay in the medium. Some questioned estimates indicate that in this scenario the Z peak width could be modified by only $\Delta \Gamma \sim 1.5$~MeV in a $T=1$~GeV medium~\cite{kapusta,contraKapusta}, which becomes experimentally negligible. 
On the other hand, in the most accepted picture of the QGP formation and evolution, at the LHC this state of matter would be 
formed after $\tau_{strong}\sim1/\Lambda_{QCD}\sim 1$~fm/c of the initial hard interaction, would be quickly thermalized in approximately $0.1$~fm/c, and might last $\sim 10$~fm/c. In this representation, weak bosons are produced and decay before the QGP is formed. It is then natural to examine the medium influence on their decays. 
Their leptonic decay channel is measured to be of $3.36\%$ ($10.8\%$) for the Z (W) while their most probable decay is into hadrons~\cite{pdg}. 
It is known that hadrons are affected by the QGP, but what happens to leptons?
Leptons interact electromagnetically with the (electric) charges of the medium. They could then suffer elastic interactions and even radiate a photon (by bremsstrahlung) after that. 
The probability of an elastic interaction to occur is determined by the lepton mean free path ($\lambda_{coll}$). Theoretical calculations suggest that $\lambda_{coll} \sim10$~fm in a $T=1$~GeV medium\footnote{
A $T=1$~GeV medium is usually considered for numerical illustrative calculations in extreme conditions.
}~\cite{blaizot}. 
Therefore, rough qualitative estimates can be done assuming that the lepton undergoes in average one elastic collision. If such, the collisional energy loss ($\Delta E_{coll}$) becomes negligible, as $\langle \Delta E_{coll} \rangle \propto \alpha_{em} \, T < 10$~MeV per elastic collision\footnote{
$\alpha_{em}$ stands for the electromagnetic coupling constant.}. 
Besides, there is a probability $\propto \alpha_{em} \, ln \big( \frac{m_D}{m}\big)$ that the lepton radiates an energy $\sim E$ after one (elastic) interaction, although in most cases it will not radiate\footnote{
The radiative energy loss after one elastic interaction is $\langle \Delta E_{rad} \rangle \propto \alpha_{em} \, E \, ln \big( \frac{m_D}{m} \big)$, where $m_D=e\,T/\sqrt{3}$ stands for the Debye mass~\cite{peigne}.
}~\cite{peigne}. Numerical calculations indicate that the typical radiative energy loss can be experimentally ignored. 
In sum, even in the most extreme conditions weak bosons and their leptonic decays behave as medium-blind references. 
\newline
As medium insensitive probes produced in initial hard parton scatterings, they could be used to validate the binary scaling (with the number of nucleon-nucleon collisions) and they could become references to observe medium induced effects on other probes, such as jets (Z-jet correlations, Sec.~\ref{sec:Zjet}) or high-$p_t$ heavy-quarks (energy loss, Sec.~\ref{sec:bcEloss}). 
\newline
\newline
Experimentally, vector bosons are measured via their leptonic decays. At the LHC, ALICE, ATLAS and CMS experiments plan to do those measurements in HIC. 
CMS~\cite{cmsHIptdr,cmsHiggsptdr} will be able to detect electrons of $p_t \ge 5$-$10$~GeV/c for $|\eta|<3$ and muons of $p_t \ge 3.5$~GeV/c for $|\eta|<2.4$. 
ALICE~\cite{alicepprI,alicepprII}  capabilities extend to electrons of $p_t >1$~GeV/c for $|\eta|<0.9$ and muons of $p_t > 1$~GeV/c for $-4< \eta < -2.5$. 
ATLAS performance studies in HIC are currently under progress, so here we do not refer to them. 
In Sec.~\ref{sec:Zbosons} Z bosons detection via the dilepton invariant mass (Sec.~\ref{sec:Dileptons}) and the Z-jet correlations (Sec.~\ref{sec:Zjet}) are discussed. Sec.~\ref{sec:Wbosons} concentrates on W bosons measurement via the single lepton spectra (Sec.~\ref{sec:SingleLepton}) and on their potential to learn on heavy-quark energy loss (Sec.~\ref{sec:bcEloss}). 

\section{Z bosons}
\label{sec:Zbosons}

The dilepton invariant mass permits to cleanly identify the Z boson and measure its properties. Both CMS and ALICE experiments plan to measure Z bosons in heavy-ion collisions. 

\subsection{The dilepton invariant mass}
\label{sec:Dileptons}

\begin{figure}[!htb]
  \centering
  \includegraphics[width=0.9\columnwidth]{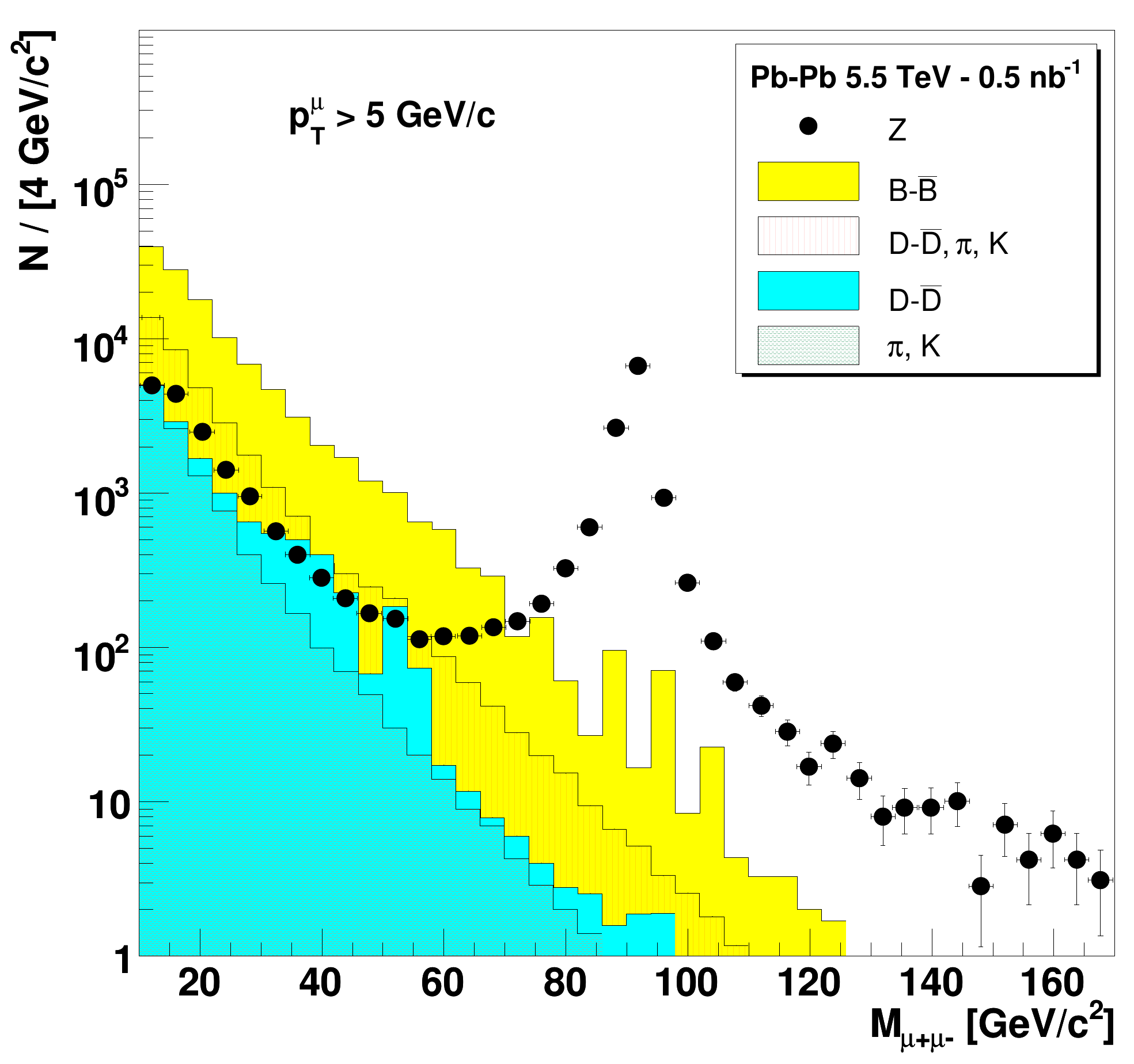}
\caption{Estimated dimuon invariant mass in Pb-Pb collisions at
  $5.5$~TeV by CMS~\cite{cmsHIptdr}. Distribution for $p_t^{\mu}>5$~GeV/c
  and $|\eta^{\mu}|<2.4$ for $\mathcal{L}=0.5$~nb$^{-1}$.}
\label{fig:dimuons-cms-Pb-Pb}
\end{figure}

\begin{figure}[!htb]
  \centering
  \includegraphics[width=0.9\columnwidth]{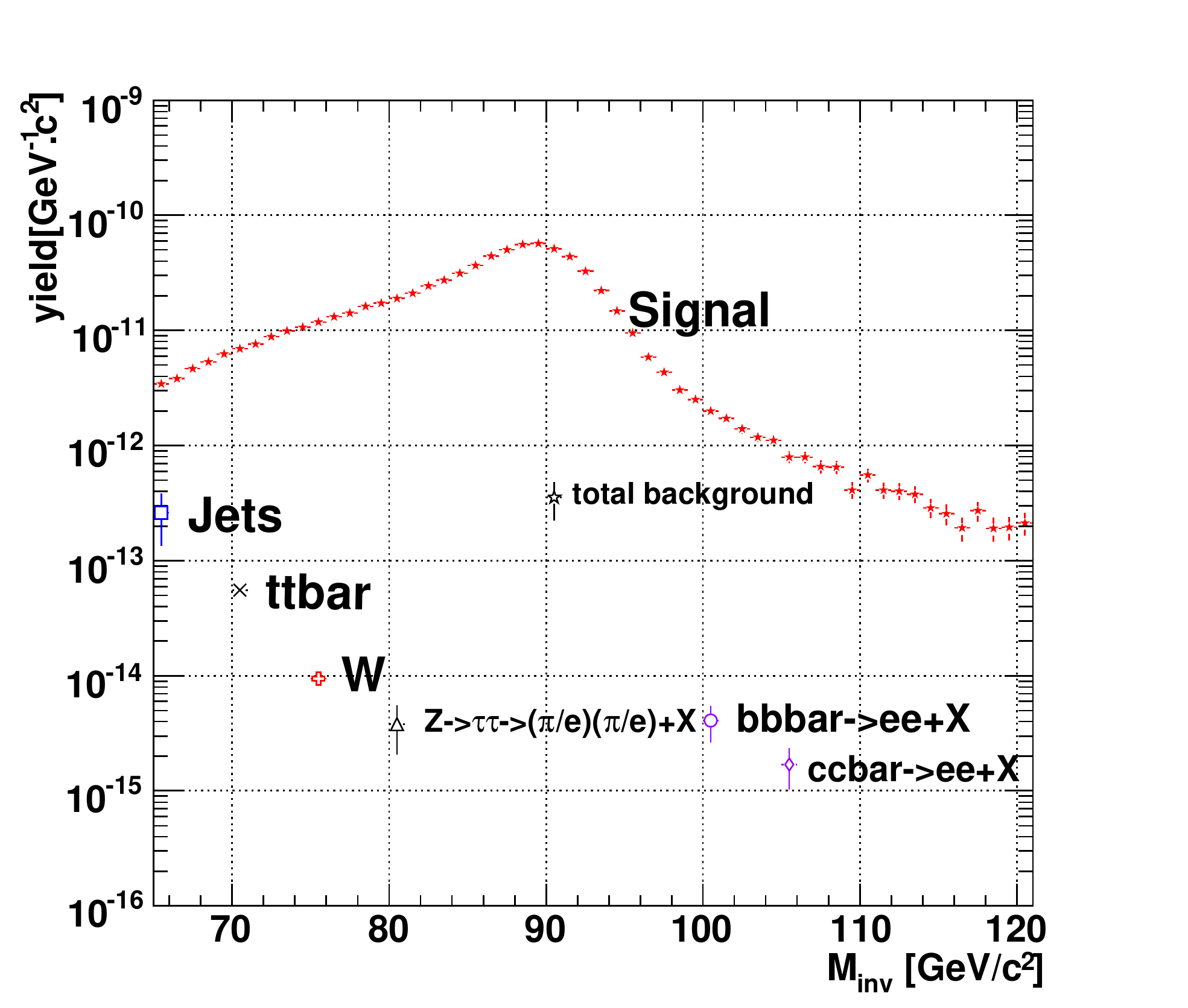}
\caption{Estimated dielectron invariant mass in pp collisions at $14$~TeV by ALICE~\cite{RaphaelleQM}. Distribution for $p_t^{e}>25$~GeV/c and $|\eta^{e}|<0.9$, applying isolation cuts. Background contributions are averaged over $66 < M_{e^+ e^-} < 116$~GeV/c$^2$ and displayed as points at an arbitrary mass value.
  }
\label{fig:dielectrons-alice-pp}
\end{figure}

CMS experiment capabilities~\cite{cmsHIptdr,cmsHiggsptdr} allow to measure the Z both in the dimuon and dielectron decay channels with similar statistics from pp to Pb-Pb collisions. For illustration, Fig.~\ref{fig:dimuons-cms-Pb-Pb} presents their expected dimuon invariant yield in Pb-Pb collisions at $5.5$~TeV. In the region of $10< M_{\mu^+ \mu^-}<70$~GeV/c$^2$ $b\bar{b}$ decays are majoritary. This region is suitable to study heavy quark energy loss and CMS should be able to exploit it by reducing the Drell-Yan contribution via b-tagging techniques. % ($\delta r > 50$~$\mu$m cut). 
At higher mass ($M$) the Z contribution prevails. The Z peak is clearly mesurable whereas the background is remarkably small (it amounts to less than $5\%$ in $M_Z \pm 5$~GeV/c$^2$). Estimations indicate that about $11 000$ Z could be reconstructed into dimuons for $M_Z \pm 10$~GeV/c$^2$ in Pb-Pb collisions at $5.5$~TeV considering an integrated luminosity of $\mathcal{L}=0.5$~nb$^{-1}$, and $\sim 10^7$ Z into dielectrons in pp collisions at $14$~TeV for $\mathcal{L}=100$~fb$^{-1}$.
\newline
ALICE capabilities turn on quite different patterns in the dielectron and dimuon decay channels. The Transition Radiation Detector and the Time Projection Chamber are the electron identification devices in ALICE, then the ability to reject the misidentified pions becomes the challenge to identify high-$p_t$ electrons. Calculations~\cite{RaphaelleQM} indicate that this can be left ahead applying track isolation criteria, considering only clean and isolated tracks (rejecting tracks which are at $|\Delta \eta| \le 0.1$, $|\Delta \phi| \le 0.1$~rad and have $p_t > 2$~GeV/c). The expected yield into dielectrons in pp collisions at $14$~TeV is shown in Fig.~\ref{fig:dielectrons-alice-pp}. 
About a few thousands Z should be reconstructed into $e^+ e^-$ \& $\mu^+ \mu^-$ in pp collisions at $14$~TeV by ALICE for $\mathcal{L}=30$~pb$^{-1}$, while trigger strategies have to be optimized for Pb-Pb collisions~\cite{RaphaelleQM,zaidaHQ,zaidaPhd}.
\newline
Apart from what has been mentioned, the Z can also be potentially used as a reference for quarkonia normalization, as though far (in terms of $M$) the Z is the only unmodified resonance.

\subsection{Z-jet correlations}
\label{sec:Zjet}

\begin{figure}[!htb]
  \centering
  \includegraphics[width=0.8\columnwidth]{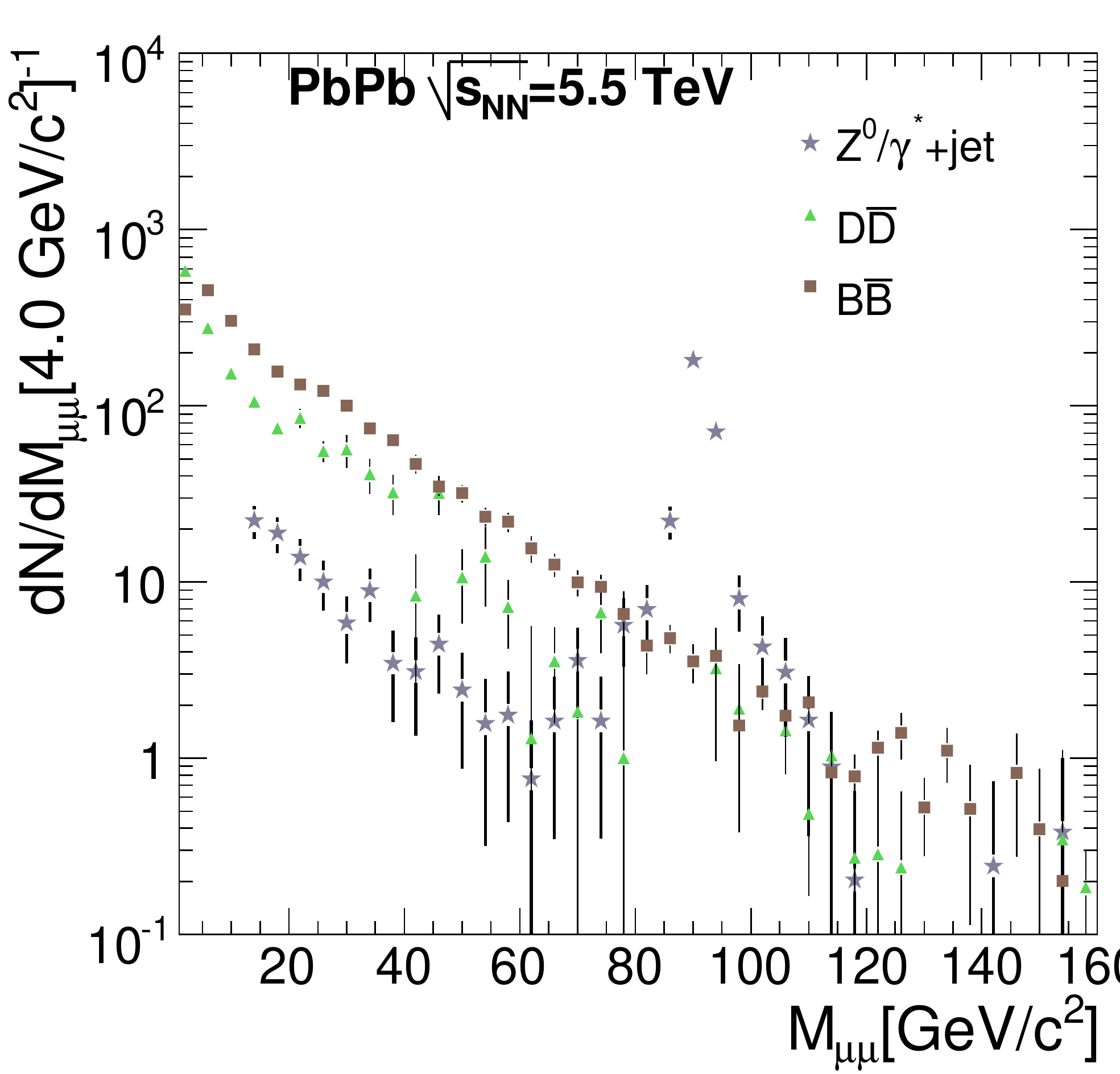}
\caption{Estimated dimuon-jet invariant mass in Pb-Pb collisions at
  $5.5$~TeV by CMS~\cite{cmsHIptdr}. Distribution for $p_t^{\mu^+ \mu^-}>25$~GeV/c,
  $p^{\mu}>3.5$~GeV/c and $|\eta^{\mu}|<2.4$ for $\mathcal{L}
  =0.5$~nb$^{-1}$.}
\label{fig:dimuons-jet-cms-Pb-Pb}
\end{figure}

In the parton model of pQCD, at LO jets are always produced by pairs, with equal transverse momenta and opposite sense of movement. The jets being influenced by the medium while the Z bosons are not, Z-jet correlations are the ideal reference for jet calibration and to study the jet fragmentation functions and their modifications in the QGP. 
The CMS experiment has demonstrated its capabilities to perform such measurements, whereas ATLAS and ALICE have still no conclusive studies. CMS expects to reconstruct a few thousand Z-jet in the dimuon decay channel for $p_t^{\mu^+ \mu^-}>25$~GeV/c and $\mathcal{L} =0.5$~nb$^{-1}$ in Pb-Pb collisions at $5.5$~TeV~\cite{cmsHIptdr} (Fig.~\ref{fig:dimuons-jet-cms-Pb-Pb}). In addition, their abilities allow to reject the charm and beauty correlated contributions by up to a factor $5$ via a $3\sigma$ cut on the vertex Distance of Closest Approach. The expected performances indicate that they will be sensitive to the jet fragmentation functions and to their modifications (see details in references~\cite{cmsHIptdr,Mironov,Kunde}).

\section{W bosons}
\label{sec:Wbosons}

W bosons production cross-section in the leptonic decay channel is ten times larger than the Z bosons (see Tab.~\ref{tab:cross_section}). They are commonly observed through their leptonic decays ($W^+ \rightarrow \mu^+ \, \nu_{\mu}$), via the computation of the invariant mass of the lepton and the reaction missing energy ($E_t$) that is associated to the neutrino. 
At the LHC, only ATLAS and CMS experiments have the capability to perform such measurements, but at the present time there is no available up-to-date calculations of their expectations for HIC. Despite of the lack of the missing $E_t$ information, the simple single lepton high-$p_t$ distribution (the decayed leptons have a mean $p_t \approx M_W/2 \sim 40$~GeV/c) can already provide valuable information, such as the integrated production cross-section.

\subsection{The single lepton spectra}
\label{sec:SingleLepton}

\begin{figure}[!htbp]
  \centering
  \includegraphics[width=1.1\columnwidth]{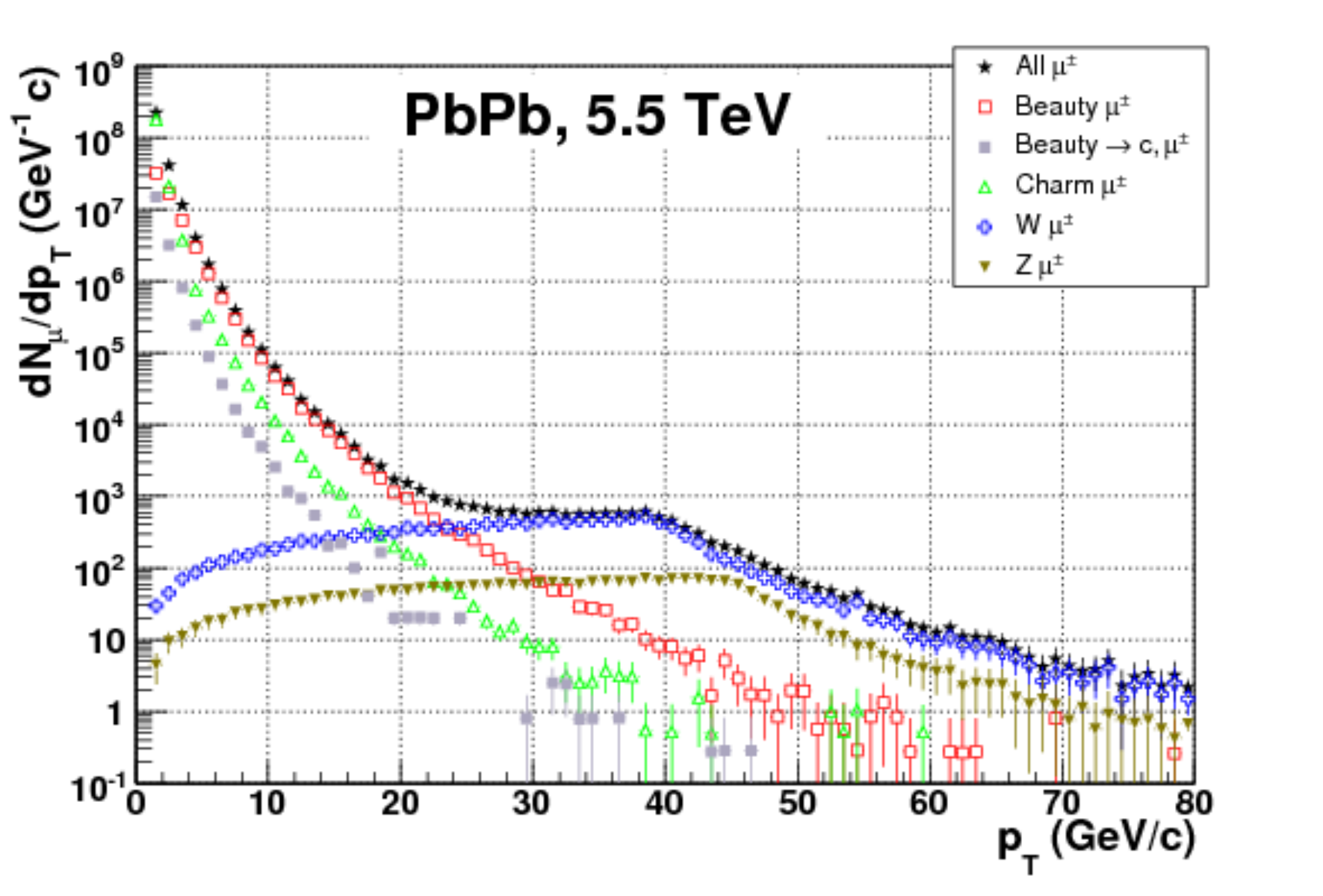}
\caption{Estimated single muon reconstructed spectra by
  ALICE for minimum bias Pb-Pb collisions at
  $5.5$~TeV. Considerations: $p_t^{\mu}>1$~GeV/c, $-4< \eta^{\mu} < -2.4$
  and $\mathcal{L} =0.5$~nb$^{-1}$ are assumed~\cite{zaidaHQ,zaidaPhd}.
 }
\label{fig:single-muon-alice-Pb-Pb}
\end{figure}

ATLAS and CMS have excellent capabilities to detect high-$p_t$ leptons, so they can perform those measurements, but the most recent studies for HIC have been done by ALICE. 
While the detection of the high-$p_t$ electron spectra by ALICE is still under study, they will be able to perform those measurements in the muonic decay channel. Fig.~\ref{fig:single-muon-alice-Pb-Pb} presents the expected reconstructed spectra for Pb-Pb collisions at $5.5$~TeV~\cite{zaidaHQ,zaidaPhd}. 
Heavy quark decays are the dominant lepton source at intermediate $p_t$, from $5$ to $25$~GeV/c, whereas W decays prevail between $30$ and $50$~GeV/c. They can then be measured via the single lepton spectra and could be used as reference to observe heavy-quark energy loss in the intermediate $p_t$ region (Sec.~\ref{sec:bcEloss}). 
The expected statistics are large enough: about $80$ ($14$) thousand reconstructed muons from $W$ decays in pp (Pb-Pb)
collisions at $14$~($5.5$)~TeV assuming $\mathcal{L} =30$~pb$^{-1}$~($0.5$~nb$^{-1}$) and no trigger downscaling~\cite{zaidaHQ,zaidaPhd}. Note that the estimates for Pb-Pb collisions will fluctuate with the trigger strategy.
\newline
A peculiar characteristic of W decays is highlighted in references~\cite{zaidaHQ,zaidaPhd}. The collision isospin induces a charge asymmetry on W production. In addition, angular momentum conservation and parity violation influence the angular distribution of W decays. For leptons of $20 < p_t < 60$~GeV/c, only W bosons cause those charge asymmetries, so
the relative abundance of positive and negative charged leptons as a function of $p_t$ and $\eta$ could be used to point out W production.

\subsection{Learning on heavy quark energy loss}
\label{sec:bcEloss}

\begin{figure}[!htbp]
  \centering
  \includegraphics[width=0.9\columnwidth]{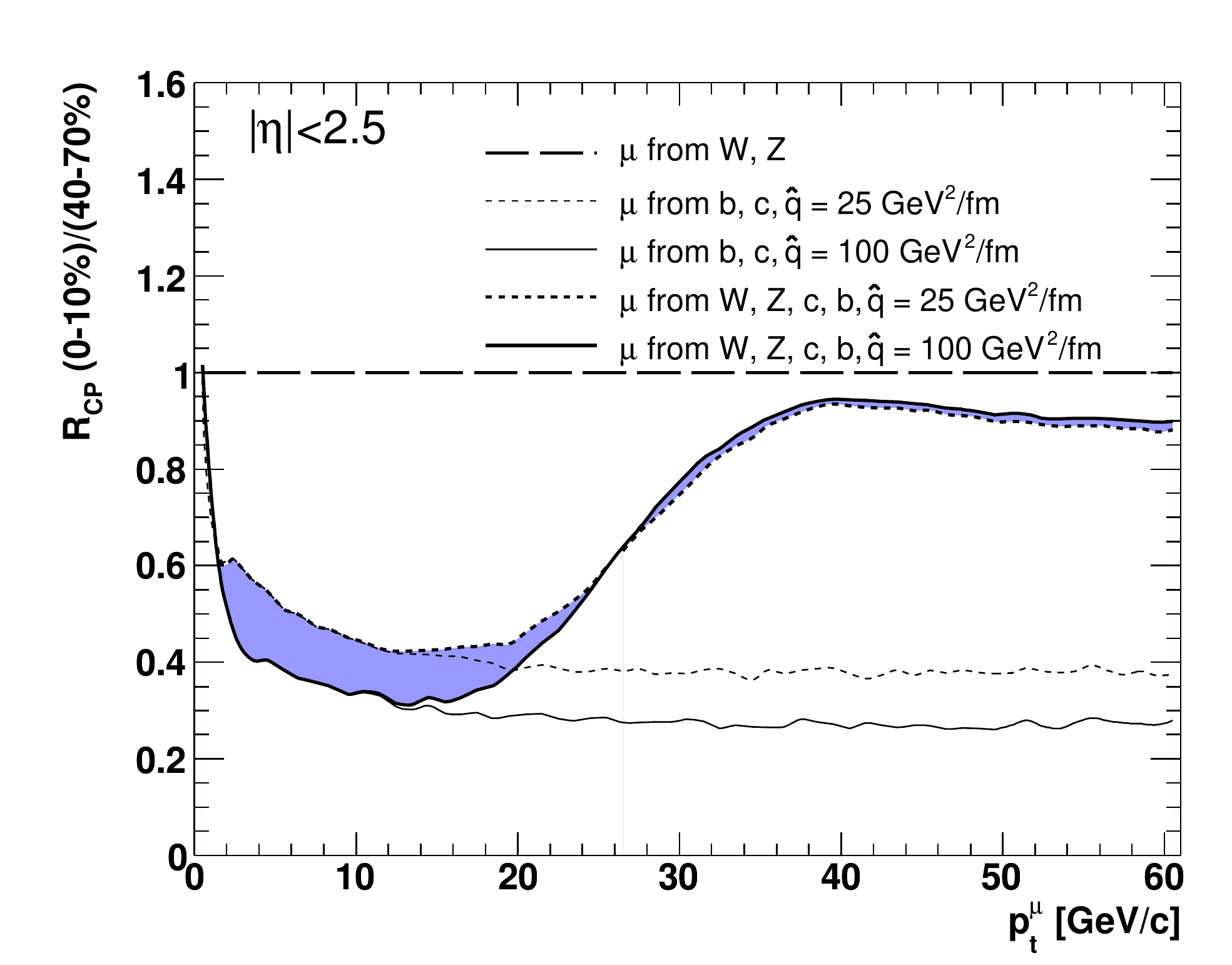}
\caption{Muon nuclear modification factor central-to-peripheral ratio, $R_{CP}(p_t)$, for $|\eta|< 2.5$~\cite{muonbcEloss}. Only heavy quarks and weak bosons contributions are included. The centralities considered are $0$-$10\%$ and $40$-$70\%$.
  }
\label{fig:Muon-Rcp-MidRapidity}
\end{figure}

In reference~\cite{muonbcEloss} the influence of heavy quarks energy loss on the single muon $p_t$ distribution is computed. The different contributions to the $p_t$ region studied are reflected in Fig.~\ref{fig:single-muon-alice-Pb-Pb}. 
Calculations are done in the framework of the Parton Quenching Model~\cite{PQM} but including an energy density evolution in pseudo-rapidity to extrapolate the results to the high-rapidity region. The authors compute the nuclear modification factor $R_{AB}(p_t)$ and the central-to-peripheral ratio $R_{CP}(p_t)$ at mid- and forward-rapidities, the regions covered by the LHC experiments. As an example, Fig.~\ref{fig:Muon-Rcp-MidRapidity} presents the $R_{CP}(p_t)$ for $|\eta| < 2.5$. 
By extrapolating the $\hat{q}$ parameter\footnote{
In this approach, the parameter $\hat{q}$ represents the time-averaged squared transferred
momentum from the medium to the parton per path length unit.
} values matching RHIC data according to the expected charged particle multiplicity, the range of values for the LHC is estimated to be $\hat{q}_{LHC}=25$-$100$~GeV$^2$/fm. 
Results lie within the shaded-band on the plot. It is obvious that the muon $p_t$ distribution is noticeably affected by
heavy-quark energy loss. The $R_{AA}$ ($R_{CP}$) muon ratios are reduced in the intermediate $p_t$ of ($2,15$)~GeV/c by a factor from $2$ to $5$ (from $2$ to $3$). The trends evolve in $p_t$ with a step rise, and for $p_t>30$~GeV/c the ratios are governed by the relative weak boson contribution. Note that the ratios shape is dependent on the rapidity window observed. In conclusion, weak bosons (W mainly) can be a reference to observe heavy-quark energy loss in the single lepton $p_t$ distribution. 

\section{Summary}
\label{sec:Summary}

Weak bosons are observed via their leptonic decays and can certainly be considered as medium blind references. 
As suggested 'standard candles', they will enable luminosity measurements and will permit to access the quarks PDFs and their nuclear modifications. Their measurement will allow to validate the Glauber binary scaling and they will be useful references to observe medium induced effects on other probes. 
Z bosons could be references for quarkonia suppression and heavy quark energy loss, while Z-jet correlations provide a clean jet calibration and the possibility to measure the jet fragmentation functions and their modifications. 
Calculations indicate that the single lepton high-$p_t$ region, predominantly populated by W decays, could be a reference to study heavy-quark energy loss in the intermediate-$p_t$ domain.
\newline
Experimentally, the LHC detectors show different and complementary capabilities to
measure and exploit weak bosons.

\begin{acknowledgement}

The author would like to thank the conference organizers for their invitation. 
Special gratitude to D. d'Enterria and G. Mart\'{\i}nez Garc\'{\i}a for fruitful discussions, and R. Granier de Cassagnac for careful reading of this manuscript. 

\end{acknowledgement}

\end{document}